\documentclass[aps,twocolumn,showpacs]{revtex4}
\usepackage{amsmath}
\usepackage{epsfig}
\usepackage{multirow}
\usepackage{appendix}
\usepackage{amssymb}
\usepackage{epstopdf}

\begin{document}
	
\title{$N\Lambda$ interaction from a quark model: $p\Lambda$ femtoscopy and the hypertriton}
\author{Ye Yan$^{1}$}\email{yanye@czcit.edu.cn}
\author{Qian Wu$^{2,3}$}
\author{Qi Huang$^4$}\email{06289@njnu.edu.cn(Corresponding author)}
\author{Hongxia Huang$^4$}\email{hxhuang@njnu.edu.cn(Corresponding author)}
\author{Jialun Ping$^4$}

\affiliation{$^1$Department of Physics, Changzhou University of Information Technology, Changzhou 213164, China}
\affiliation{$^2$School of Physics, Nanjing University, Nanjing 21000, China}
\affiliation{$^3$Institute of Modern Physics, Chinese Academy of Sciences, Lanzhou 730000, China}
\affiliation{$^4$Department of Physics, Nanjing Normal University, Nanjing 210023, China}

\begin{abstract}
	
Understanding the low-energy nucleon--$\Lambda$ interaction is essential for describing hypernuclear structures, while its spin dependence remains insufficiently constrained by experimental data. 
In this work, we investigate the $N\Lambda$ interaction using the quark delocalization color screening model and explore its implications for $p\Lambda$ femtoscopy and the hypertriton. 
The energy and scattering phase shifts of the $N\Lambda$ system are calculated, revealing weak attraction in both the $J^P=0^+$ and $1^+$ channels, but no bound states. 
The corresponding scattering lengths and effective ranges are found to be $-2.36$ fm and $1.61$ fm for $J^P=0^+$, and $-0.95$ fm and $1.46$ fm for $J^P=1^+$, respectively.
Using the Gel'fand--Levitan--Marchenko inverse-scattering method, effective $N\Lambda$ potentials are reconstructed from the coupled-channel scattering phase shifts and applied to calculate the $p\Lambda$ femtoscopic correlation function. 
Comparison with experimental data suggests that the spin-triplet attraction may be underestimated in the present model. 
Furthermore, employing the Gaussian Expansion Method, we investigate the $NN\Lambda$ three-body system and obtain a weakly bound hypertriton with a $\Lambda$ separation energy of $1.06$--$1.16$ MeV. 
These results demonstrate that $p\Lambda$ femtoscopic correlations and hypertriton binding provide complementary probes of the spin-dependent $N\Lambda$ interaction, offering valuable constraints on its low-energy behavior.

\end{abstract}
	

\maketitle

\setcounter{totalnumber}{5}

\section{Introduction}

Understanding hadron--hadron interactions is a cornerstone of nuclear and particle physics, as it provides crucial insights into the nonperturbative dynamics of Quantum Chromodynamics (QCD) and helps reveal the properties of strongly interacting matter~\cite{Fabbietti:2020bfg,Povh:1981nz,Hyodo:2020czb,Frankfurt:2022jns}. 
In particular, the nucleon-hyperon ($NY$) interaction, especially the $N \Lambda$ interaction, serves as a critical probe of the baryon--baryon force beyond the nucleon--nucleon sector~\cite{Haidenbauer:2013oca,Rijken:2006ep,Fujiwara:2006yh}. 
Its study is essential for interpreting hypernuclear properties~\cite{Dalitz:1958zza,Haidenbauer:2019boi,Petschauer:2020urh,Chen:2025eeb,Hashimoto:2006aw,Gal:2016boi,ALICE:2019vlx,STAR:2019wjm}, constraining the equation of state of dense astrophysical environments such as neutron stars~\cite{Vidana:2013nxa,Chatterjee:2015pua,Gal:2016boi,Tolos:2020aln,Gerstung:2020ktv,Tong:2025sui}, and exploring possible exotic states~\cite{Guo:2017jvc,Chen:2022asf}. 
Recent advances in scattering experiments~\cite{CLAS:2021gur,BESIII:2024geh} and femtoscopy~\cite{ALICE:2021njx} further underscore the need for the determination of the $N \Lambda$ interaction, which remains less constrained than its $NN$ counterpart due to insufficient experimental data.

Experimental studies of $\Lambda$--proton scattering have been conducted over several decades~\cite{Charlton:1970bv,Anderson:1975rh,Mount:1975pb,Hauptman:1977hr,Bassano:1967kbh,Sechi-Zorn:1968mao,Alexander:1968acu,Kadyk:1971tc,CLAS:2021gur,BESIII:2024geh,ALICE:2021njx}, spanning a wide range of energies from below 100 MeV/$c$ to over 20 GeV/$c$.
Early bubble chamber experiments in the 1960s and 1970s~\cite{Bassano:1967kbh,Sechi-Zorn:1968mao,Alexander:1968acu} provided the first measurements of elastic and inelastic cross sections, angular distributions, and polarization effects, primarily in the low- to medium-momentum regime. 
These studies established that $\Lambda p$ scattering is dominated by $S$-wave interactions at low energies, with evidence of $P$- and $D$-wave contributions emerging at higher momenta~\cite{Kadyk:1971tc}. 
In addition, the COSY-TOF Collaboration have provided complementary constraints on low-energy $\Lambda p$ scattering through the analysis of final-state interactions in the $\vec{p}p \rightarrow pK^{+}\Lambda$ reaction. 
Using polarized proton beams, the measurements disentangled the spin dependence of the interaction and yielded a model-independent extraction of the spin-triplet scattering length~\cite{COSY-TOF:2013uqx,COSY-TOF:2016qxd}.
More recent experiments, such as those conducted with the CLAS detector~\cite{CLAS:2021gur}, have significantly improved the precision and energy coverage of $\Lambda p$ scattering data, particularly in the 0.9--2.0 GeV/$c$ range, providing valuable constraints on theoretical descriptions of the $\Lambda p$ interaction, with implications for hypernuclear physics and dense matter.
Very recently, the BESIII Collaboration reported measurements of $\Lambda p \rightarrow \Lambda p$ scattering as well as the first observation of antihyperon--nucleon scattering, $\bar{\Lambda}p \rightarrow \bar{\Lambda}p$, including differential cross sections~\cite{BESIII:2024geh}. 
These results open a new frontier in the experimental exploration of hyperon--nucleon interactions~\cite{Wang:2024whi}.

While the short lifetime of hyperons makes conventional scattering experiments challenging, early measurements nevertheless provided the first constraints on the $\Lambda p$ interaction strength.
In recent years, femtoscopic correlation functions have emerged as a powerful technique, enabling access to low-energy scattering parameters~\cite{Lednicky:1981su,Lednicky:2005tb,Liu:2024uxn} and emission source properties~\cite{ALICE:2020ibs,ALICE:2023sjd,Mihaylov:2023pyl,VazquezDoce:2024nye,Wang:2025htd,ALICE:2025zzg,ALICE:2025,Wang:2024bpl} through the analysis of final-state momentum correlations in heavy-ion and proton--proton collisions.
Femtoscopic correlation functions have been extensively studied in recent years, covering baryon--baryon~\cite{ALICE:2020mfd,STAR:2015kha,STAR:2014dcy,ALICE:2019hdt,Morita:2014kza,Ohnishi:2016elb,Hatsuda:2017uxk,Haidenbauer:2018jvl,Ohnishi:2021ger,Kamiya:2021hdb,Zeng:2025kur,Si:2025eou,Tang:2025bcc,Wang:2022nxr,Huang:2020iil,Gobel:2025afq}, baryon--meson~\cite{ALICE:2019gcn,ALICE:2021cpv,ALICE:2022enj,Kamiya:2019uiw,Haidenbauer:2021zvr,Liu:2022nec,Sarti:2023wlg,Molina:2023jov,Feijoo:2024bvn,Abreu:2024qqo,Li:2024tvo,Encarnacion:2025luc,Agatao:2025ckp,Barbat:2025orm,Liu:2025eqw,Bahmani:2020jjq}, meson--meson~\cite{ALICE:2018nnl,Graczykowski:2021vki,Liu:2023uly,Molina:2023oeu,Vidana:2023olz,Albaladejo:2023wmv,Feijoo:2023sfe,Li:2024tof,Liu:2024nac,Ikeno:2023ojl,Torres-Rincon:2023qll,Shen:2025qpj,Albaladejo:2025lhn,Abreu:2025jqy,Encarnacion:2025lyf}, baryon--nucleus~\cite{STAR:2024zvj,Mrowczynski:2025qys,Ogata:2021mbo,Etminan:2024nak,Etminan:2024ukl}, nucleus-nucleus~\cite{Mrowczynski:2021bzy,Ge:2025put,Qiao:2024wha,Wang:2023ygv}, and three-body systems~\cite{ALICE:2025aur,ALICE:2022boj,Garrido:2025lar}.
In particular, studies of $p\Lambda$ correlations have evolved from source-size measurements in Au+Au collisions to high-precision investigations at the LHC~\cite{Chung:2002vk,NA49:2002rmy,ALICE:2018ysd,STAR:2005rpl}.
More recently, the ALICE Collaboration has measured the $p\Lambda$ correlation function in $pp$ collisions with high precision~\cite{ALICE:2021njx}, revealing signatures associated with the $N\Lambda$--$N\Sigma$ coupled-channel effect and providing new constraints for theoretical models.

Significant theoretical progress has also been made in the study of the $p\Lambda$ correlation function.
In particular, Ref.~\cite{Wang:1999bf} explored the feasibility of utilizing $p\Lambda$ correlations to probe the source size in relativistic heavy-ion collisions.
The impact of residual correlations on both $p\Lambda$ and $p\bar{\Lambda}$ systems was systematically addressed in Ref.~\cite{Shapoval:2015sba}.
Moreover, Refs.~\cite{Shapoval:2014yha,Mihaylov:2023ahn} focused on extracting the low-energy scattering parameters from the $p\Lambda$ correlation function.
More recently, Ref.~\cite{Garrido:2024pwi} extended these efforts by investigating the $pp\Lambda$ correlation function on the basis of the $p\Lambda$ correlations.
Together, these studies have gradually established the $p\Lambda$ correlation function as a powerful tool for probing both source dynamics and $\Lambda$--nucleon interactions.

A wide range of theoretical frameworks has been employed to investigate the $N\Lambda$ interaction.
Constituent quark models, such as the fss2, have been extended to the $YN$ sector and have provided valuable insight into the short-range dynamics~\cite{Zhang:1997ny,Fujiwara:2004jj,Fujiwara:2006yh}. 
Descriptions of the $N\Lambda$ interaction have also been developed within meson-exchange frameworks, particularly by the Jülich and Nijmegen groups~\cite{Maessen:1989sx,Rijken:1998yy,Tominaga:2001ra,Nagels:2015lfa,Holzenkamp:1989tq,Reuber:1992dh,Haidenbauer:2005zh}. 
In addition, chiral effective field theory (EFT) provides a systematically improvable framework for describing $N\Lambda$ interactions, with developments up to next-to-next-to-leading order~\cite{Polinder:2006zh,Haidenbauer:2013oca,Li:2016paq,Li:2016mln,Haidenbauer:2019boi,Song:2021yab,Haidenbauer:2023qhf}.

In this work, we employ the quark delocalization color screening model (QDCSM) to investigate the $N\Lambda$ interaction and its implications for different observables. 
Specifically, we first calculate the $N\Lambda$ scattering phase shifts and extract the corresponding scattering parameters. 
Using the Gel’fand--Levitan--Marchenko (GLM) inverse scattering method, we reconstruct effective $N\Lambda$ potentials from the coupled-channel scattering information. 
On this basis, the $p\Lambda$ femtoscopic correlation function is calculated, taking into account spin averaging and feed-down contributions. 
The $N\Lambda$ interaction is also particularly important for the structure of the hypertriton ${}^3_\Lambda\mathrm{H}$~\cite{Bertulani:2022vad,Hildenbrand:2019sgp,Schafer:2021qbt,Gajewski:1967ruj,Juric:1973zq,ALICE:2022sco}, whose very small $\Lambda$ separation energy makes it a sensitive probe of low-energy $N\Lambda$ dynamics. 
Therefore, the Gaussian Expansion Method (GEM) is employed to solve the $NN\Lambda$ three-body Schrödinger equation and investigate the binding of ${}^3_\Lambda\mathrm{H}$.

The structure of this paper is as follows:
In Sec.~\ref{2}, we present the theoretical framework, including the QDCSM, the formalism for the two-particle correlation function, and the GLM inverse scattering method.
Numerical results and comparisons are discussed in Sec.~\ref{3}, followed by a summary.

\section{THEORETICAL FORMALISM}
\label{2}

\subsection{Quark delocalization color screening model}
\label{21}

The QDCSM is an extension of the quark cluster model~\cite{Wang:1992wi} and has been widely applied to the study of multiquark systems, such as $\Delta\Delta$~\cite{Huang:2013nba} and $N\Omega$ dibaryons~\cite{Yan:2024aap}.
In this section, we briefly introduce the main ingredients of the QDCSM relevant to the present calculation.

The Hamiltonian of a dibaryon system is written as
\begin{align}
	H = \sum_{i=1}^{6} \left( m_i+\frac{\boldsymbol{p}_i^{\,2}}{2m_i} \right) -T_{\mathrm{c.m.}} +\sum_{i<j}^{6}V(\boldsymbol{r}_{ij}),
\end{align}
where $m_i$ and $\boldsymbol{p}_i$ are the mass and momentum of the $i$th quark, respectively, and $T_{\mathrm{c.m.}}$ denotes the center-of-mass kinetic energy. 
The interaction between two constituent quarks is composed of the confinement, one-gluon-exchange, and Goldstone-boson-exchange interactions,
\begin{align}
	V(\boldsymbol{r}_{ij}) = V_{\mathrm{conf}}(\boldsymbol{r}_{ij}) + V_{\mathrm{OGE}}(\boldsymbol{r}_{ij}) + V_{\mathrm{OBE}}(\boldsymbol{r}_{ij}).
\end{align}

A characteristic feature of the QDCSM is the phenomenological color-screening confinement interaction,
\begin{align}
	V_{\mathrm{conf}}(\boldsymbol{r}_{ij}) = -a_c \boldsymbol{\lambda}_i^{c}\cdot\boldsymbol{\lambda}_j^{c}\left[f(\boldsymbol{r}_{ij})+V_0\right],
\end{align}
where
\begin{align}
	f(\boldsymbol{r}_{ij})=
	\begin{cases}
		r_{ij}^{2},
		& i,j \text{ in the same cluster}, \\[2mm]
		\dfrac{1-\exp\left(-\mu_{q_iq_j}r_{ij}^{2}\right)}
		{\mu_{q_iq_j}},
		& i,j \text{ in different clusters}.
	\end{cases}
\end{align}
Here, $a_c$, $V_0$, and $\mu_{q_iq_j}$ are model parameters, and
$\boldsymbol{\lambda}^{c}$ denotes the SU(3) color Gell-Mann matrices.
The color-screening parameters are determined by reproducing the deuteron properties as well as nucleon-nucleon and hyperon-nucleon scattering phase shifts.

In the present work, we focus on the low-lying positive-parity
$S$-wave $N\Lambda$ system. Therefore, the spin-orbit and tensor interactions are omitted.
The one-gluon-exchange potential, including the color-Coulomb and chromomagnetic interactions, is written as
\begin{align}
	V_{\mathrm{OGE}}(\boldsymbol{r}_{ij}) =& \frac{1}{4}\alpha_s(\mu) \boldsymbol{\lambda}_i^{c} \cdot \boldsymbol{\lambda}_j^{c} \nonumber\\
	&\times \left[ \frac{1}{r_{ij}} -\frac{\pi}{2} \delta^{(3)}(\boldsymbol{r}_{ij}) \left( \frac{1}{m_i^{2}} +\frac{1}{m_j^{2}} +\frac{4\boldsymbol{\sigma}_i\cdot\boldsymbol{\sigma}_j} {3m_i m_j} \right) \right],
	\label{Voge}
\end{align}
where $\boldsymbol{\sigma}$ denotes the Pauli spin matrices and $\alpha_s(\mu)$ is an effective scale-dependent quark-gluon coupling constant.
To account for the different quark-mass scales, $\alpha_s(\mu)$ is parametrized as~\cite{Vijande:2004he}
\begin{align}
	\alpha_s(\mu) = \frac{\alpha_0} {\ln\left[ \left(\mu^{2}+\mu_0^{2}\right)/\Lambda_0^{2} \right]},
\end{align}
where $\mu$ is the reduced mass of the interacting quark pair.

Dynamical chiral symmetry breaking gives rise to Goldstone-boson-exchange interactions among the constituent $u$, $d$, and $s$ quarks.
The corresponding interaction is expressed as
\begin{align}
	V_{\mathrm{OBE}}(\boldsymbol{r}_{ij}) =& V^{\pi}(\boldsymbol{r}_{ij}) \sum_{a=1}^{3} \lambda_i^{a}\lambda_j^{a} + V^{K}(\boldsymbol{r}_{ij}) \sum_{a=4}^{7} \lambda_i^{a}\lambda_j^{a} \nonumber\\
	&+ V^{\eta}(\boldsymbol{r}_{ij}) \left[ \lambda_i^{8}\lambda_j^{8}\cos\theta_p - \lambda_i^{0}\lambda_j^{0}\sin\theta_p \right],
\end{align}
where $\lambda^{a}$ denotes the SU(3) flavor Gell-Mann matrices and $\theta_p$ is the pseudoscalar mixing angle associated with the physical $\eta$ meson.

The potentials $V^{\pi}(\boldsymbol{r}_{ij})$, $V^{K}(\boldsymbol{r}_{ij})$, and $V^{\eta}(\boldsymbol{r}_{ij})$ share the common form, which is given below for $\chi = \pi, K, \eta$
\begin{align}
	V^{\chi}(\boldsymbol{r}_{ij}) =& \frac{g_{\mathrm{ch}}^{2}}{4\pi} \frac{m_{\chi}^{2}}{12m_i m_j} \frac{\Lambda_{\chi}^{2}} {\Lambda_{\chi}^{2}-m_{\chi}^{2}} m_{\chi} \nonumber\\
	&\times 	\left[ Y(m_{\chi}r_{ij}) - \frac{\Lambda_{\chi}^{3}}{m_{\chi}^{3}} Y(\Lambda_{\chi}r_{ij}) \right] (\boldsymbol{\sigma}_i\cdot\boldsymbol{\sigma}_j),
\end{align}
where $Y(x)=\text{e}^{-x}/x$ is the standard Yukawa function.
The meson masses $m_{\pi}$, $m_K$, and $m_{\eta}$ are taken from experiment.
The chiral coupling constant $g_{\mathrm{ch}}$ is determined from the $\pi NN$ coupling constant through
\begin{align}
	\frac{g_{\mathrm{ch}}^{2}}{4\pi} = \left(\frac{3}{5}\right)^{2} \frac{g_{\pi NN}^{2}}{4\pi} \frac{m_{u,d}^{2}}{m_N^{2}}.
\end{align}
Flavor SU(3) symmetry is assumed for the chiral interaction, with explicit symmetry breaking arising from the different strange-quark mass.

The model parameters used in the present calculation are listed in Table~\ref{parameters}.
They are fixed by reproducing the masses of the ground-state hadrons~\cite{ParticleDataGroup:2026aaa}.
The calculated masses of the ground-state baryons relevant to the present study are compared with the experimental values in Table~\ref{hadrons}.

\begin{table}[ht]
	\caption{\label{parameters}Model parameters used in this work:
		$m_{\pi} = 0.7$, $m_{K} = 2.51$, $m_{\eta} = 2.77$, $\Lambda_{\pi} = 4.2$, $\Lambda_{K} = 5.2$, $\Lambda_{\eta} = 5.2$ fm$^{-1}$, $g_\text{ch}^2/(4\pi)$ = 0.54.}
	\begin{tabular}{cccc} \hline\hline
		~~~~~$b$~~~~~ & ~~~~~$m_{q}$~~~~~ & ~~~~~~$m_{s}$~~~~~~  & ~~~$ a_c$~~~ \\
		(fm)        & (MeV)       & (MeV)          & (MeV $\cdot$ fm$^{-2}$) \\
		0.518       & 313         & 573            &  58.03                  \\ \hline
		$V_0$       & $\alpha_0$  & $\Lambda_0$    & $\mu_0$                 \\
		(MeV)       &             & (fm$^{-1}$)    & (MeV)                   \\
		$-$1.2883   & 0.510       &  1.525         & 445.808                 \\ \hline\hline	
	\end{tabular}
\end{table}

\begin{table}[ht]
	\caption{Masses of ground-state baryons (in MeV).}
	\begin{tabular}{c c c c}
		\hline \hline
		Baryon     & ~~~~~$I(J^P)$~~~~~  & ~~~~~$M^{\text{Exp}}$~~~~~ & ~$M^{\text{Theo}}$~  \\ \hline
		$N$        & $1/2(1/2^+)$ & 939   & 939    \\
		$\Delta$   & $3/2(3/2^+)$ & 1232  & 1232   \\
		$\Lambda$  & $0(1/2^+)$   & 1116  & 1124   \\
		$\Sigma$   & $1(1/2^+)$   & 1193  & 1238   \\
		$\Sigma^*$ & $1(3/2^+)$   & 1385  & 1360   \\
		$\Xi$      & $1/2(1/2^+)$ & 1318  & 1374   \\
		$\Xi^*$    & $1/2(3/2^+)$ & 1533  & 1496   \\
		$\Omega$   & $0(3/2^+)$   & 1672  & 1642   \\
		
		\hline\hline
	\end{tabular}
	\label{hadrons}
\end{table}

Another essential ingredient of the QDCSM is quark delocalization.
It enlarges the variational space of the multiquark system and allows the internal quark orbitals to adjust dynamically during the interaction.
The single-particle orbital wave functions are constructed as linear combinations of left- and right-centered Gaussian functions,
\begin{eqnarray}
	\psi_{\alpha}(\boldsymbol{S}_i,\epsilon) &=& \frac{ \phi_{\alpha}(\boldsymbol{S}_i) + \epsilon\phi_{\alpha}(-\boldsymbol{S}_i)} {N(\epsilon)}, \nonumber\\
	\psi_{\beta}(-\boldsymbol{S}_i,\epsilon) &=& \frac{ \phi_{\beta}(-\boldsymbol{S}_i) + \epsilon\phi_{\beta}(\boldsymbol{S}_i)} {N(\epsilon)}, \nonumber\\
	N(S_i,\epsilon) &=& \sqrt{ 1+\epsilon^{2} +2\epsilon \exp\left(-S_i^{2}/4b^{2}\right) }.
	\label{1q}
\end{eqnarray}
Here, $\boldsymbol{S}_i$ denotes the generator coordinate and $b$ is the Gaussian width parameter.
The delocalization parameter $\epsilon$ is not an additional fitted parameter; instead, it is determined variationally for each configuration by the dynamics of the multiquark system.
This allows the quark orbitals to rearrange themselves dynamically as the two clusters approach each other.
In addition, the baryon wave functions are constructed based on group-theoretical considerations, and their forms can be found in Ref.~\cite{Yan:2024usf}.

\subsection{Two-particle correlation function}
\label{22}

Experimentally, the two-particle correlation function can be constructed as
\begin{align}
	C(\boldsymbol{k})
	=
	\xi(\boldsymbol{k})
	\frac{N_{\mathrm{same}}(\boldsymbol{k})}
	{N_{\mathrm{mixed}}(\boldsymbol{k})},
\end{align}
where $N_{\mathrm{same}}(\boldsymbol{k})$ and $N_{\mathrm{mixed}}(\boldsymbol{k})$ denote the relative-momentum distributions of hadron pairs constructed from the same and mixed events, respectively, and $\xi(\boldsymbol{k})$ accounts for the normalization and experimental corrections.

Theoretically, the correlation function can be evaluated within the Koonin--Pratt (KP) formalism~\cite{Koonin:1977fh,Pratt:1990zq,Bauer:1992ffu},
\begin{align}
	C(\boldsymbol{k}) &= \frac{N_{12}(\boldsymbol{p}_1,\boldsymbol{p}_2)}{N_1(\boldsymbol{p}_1)N_2(\boldsymbol{p}_2)}\nonumber\\
	&\simeq\frac{ \int \text{d}^4x_1\,\text{d}^4x_2\, S_1(x_1,\boldsymbol{p}_1) S_2(x_2,\boldsymbol{p}_2) |\Psi(\boldsymbol{r},\boldsymbol{k})|^2}{\int \text{d}^4x_1\,\text{d}^4x_2\,S_1(x_1,\boldsymbol{p}_1)S_2(x_2,\boldsymbol{p}_2)}\nonumber\\
	&\simeq\int \text{d}^3r\,S_{12}(r)|\Psi(\boldsymbol{r},\boldsymbol{k})|^2 ,
	\label{KP}
\end{align}
where $S_i(x_i,\boldsymbol{p}_i)$ ($i=1,2$) is the single-particle emission source function for hadron $i$ with momentum $\boldsymbol{p}_i$.
The relative momentum is defined as
\begin{align}
	\boldsymbol{k} = \frac{ m_2\boldsymbol{p}_1-m_1\boldsymbol{p}_2 }{ m_1+m_2},
\end{align}
and $\boldsymbol{r}$ denotes the relative separation of the two particles at emission, including the correction associated with their emission-time
difference. 
The function $\Psi(\boldsymbol{r},\boldsymbol{k})$ is the relative two-particle scattering wave function with asymptotic relative momentum $\boldsymbol{k}$.

Neglecting the emission-time difference and the momentum dependence of the source, and integrating over the center-of-mass coordinate, the KP formula reduces to the last line of Eq.~(\ref{KP}). 
In the present work, a spherically symmetric Gaussian source is adopted,
\begin{align}
	S_{12}(r) = \frac{1}{(4\pi R^2)^{3/2}} \exp\left(-\frac{r^2}{4R^2}\right),
	\label{source}
\end{align}
where $R$ is the source-size parameter and $\int \text{d}^3r\,S_{12}(r)=1$.
Therefore, the correlation function contains two essential ingredients: the emission source, characterized by $S_{12}(r)$, and the final-state interaction between the two particles, encoded in $\Psi(\boldsymbol{r},\boldsymbol{k})$.

For a pair of nonidentical particles such as $p\Lambda$, and assuming that only the $S$-wave component is modified by the strong interaction, the relative wave function can be written as
\begin{align}
	\Psi_{p\Lambda}(\boldsymbol{r},\boldsymbol{k}) = \text{e}^{i\boldsymbol{k}\cdot\boldsymbol{r}} -j_0(kr) +\psi_{p\Lambda}(r,k),
	\label{wave}
\end{align}
where $j_0(kr)$ is the $S$-wave component of the noninteracting plane wave, while $\psi_{p\Lambda}(r,k)$ denotes the interacting $S$-wave component.
Substituting Eq.~(\ref{wave}) into Eq.~(\ref{KP}) and performing the angular integration gives
\begin{align}
	C_{p\Lambda}(k) = 1+ \int_{0}^{\infty} 4\pi r^2\,\text{d}r S_{12}(r) \left[ |\psi_{p\Lambda}(r,k)|^2 - |j_0(kr)|^2 \right].
	\label{Ck}
\end{align}

The interacting $S$-wave component $\psi_{p\Lambda}(r,k)$ is obtained by solving the Schr\"odinger equation with the effective $p\Lambda$ interaction.
Introducing the reduced radial wave function
\begin{align}
	u_k(r)=r\,\psi_{p\Lambda}(r,k),
\end{align}
the corresponding radial Schr\"odinger equation reads
\begin{align}
	-\frac{\hbar^2}{2\mu} \frac{\text{d}^2u_k(r)}{\text{d}r^2} + V(r)u_k(r) = E u_k(r),
	\label{eq}
\end{align}
where
\begin{align}
	\mu = \frac{m_p m_{\Lambda}} {m_p+m_{\Lambda}}, \qquad E=\frac{\hbar^2k^2}{2\mu}.
\end{align}
The radial wave function is normalized to the scattering asymptotic form such that, in the absence of the interaction,
$\psi_{p\Lambda}(r,k)$ reduces to $j_0(kr)$.
The correlation function for each spin channel is then calculated from Eq.~(\ref{Ck}).

This coordinate-space approach, in which the scattering wave function is obtained by solving the Schr\"odinger equation, is also implemented in the
femtoscopic correlation analysis tool CATS~\cite{Mihaylov:2018rva} and has been widely employed in femtoscopic studies of hadronic systems~\cite{Morita:2014kza,Ohnishi:2021ger}.
Alternatively, the scattering wave function may be obtained in momentum space by solving the Lippmann--Schwinger or Bethe--Salpeter equation~\cite{Haidenbauer:2018jvl,Liu:2023uly}.

Since the experimentally measured $p\Lambda$ correlation function is not spin resolved, the theoretical correlation function is averaged over the spin-singlet and spin-triplet channels according to their spin multiplicities,
\begin{align}
	C_{p\Lambda}(k) = \frac{1}{4}C_{p\Lambda}^{J=0}(k) + \frac{3}{4}C_{p\Lambda}^{J=1}(k).
	\label{average}
\end{align}

\subsection{Construction of the effective $N\Lambda$ potential}
\label{23}

To calculate the correlation function through the Schr\"odinger equation, an effective two-body $N\Lambda$ interaction potential $V(r)$ is required.
Since the QDCSM is formulated at the quark level for few-body systems, it is not straightforward to directly extract a local hadron--hadron potential from the microscopic Hamiltonian.
Nevertheless, the QDCSM can be applied to hadron--hadron scattering processes, from which the corresponding scattering information can be obtained and subsequently used to construct an effective two-body potential.

In the present work, the effective $N\Lambda$ potential is reconstructed from the scattering information obtained in the QDCSM by using the GLM inverse-scattering method~\cite{Chadan1977}.
In particular, the $N\Lambda$ scattering phase shifts are calculated with channel coupling included, so that the coupling effects are encoded in the scattering information and subsequently incorporated into the reconstructed effective interaction.

For the $S$ wave, the central equation of the GLM method is the Marchenko equation~\cite{Marchenko1955,Agranovich1963},
\begin{align}
	K(r,r')+F(r,r')+\int_r^\infty K(r,s)F(s,r')\,\text{d}s = 0 .
\end{align}
Here, $K(r,r')$ is the kernel function to be determined, while $F(r,r')$ is constructed from the scattering information through
\begin{align}
	F(r,r')=&\frac{1}{2\pi}\int_{-\infty}^{\infty}\text{e}^{\text{i}kr}[1-S(k)]\text{e}^{\text{i}kr'}\text{d}k \nonumber \\
	&+\sum_{i=1}^{n}M_i \text{e}^{-\kappa_i r}\text{e}^{-\kappa_i r'} .
\end{align}

The partial-wave scattering matrix is expressed as $S(k)=\exp[2\text{i}\delta(k)]$, where $\delta(k)$ is the scattering phase shift.
At low energies, the scattering phase shift can be characterized by the effective-range expansion $k\cot\delta=-1/a_0+ 1/2 r_{\text{eff}}k^2$, where $a_0$ and $r_{\rm eff}$ denote the scattering length and effective range, respectively.
When bound states are present, $n$ is the number of bound states, $\kappa_i$ denotes the wave number of the $i$th bound state, and $M_i$ is the corresponding norming constant.

After solving the Marchenko equation and obtaining $K(r,r')$, the effective potential is reconstructed as
\begin{align}
	V(r)=-2\frac{\text{d}}{\text{d}r}K(r,r).
\end{align}

The resulting potential reproduces the scattering information used as input and provides an effective coordinate-space representation of the $N\Lambda$ interaction.
In the present calculation, the effects of channel coupling contained in the QDCSM scattering phase shifts are therefore effectively incorporated into the reconstructed single-channel $N\Lambda$ potential.
The obtained potential is subsequently employed to determine the $p\Lambda$ scattering wave function entering the correlation function and is also used as the $N\Lambda$ interaction in the $NN\Lambda$ three-body calculation.
Further details of the GLM inverse-scattering formalism and its applications can be found in Refs.~\cite{Chadan1977,Marchenko1955,Agranovich1963,Newton2002,Meoto:2019jky,Khokhlov:2021afy}.

\subsection{Gaussian Expansion Method}
\label{24}

\begin{figure}[htb]
	\centering
	\includegraphics[width=7cm]{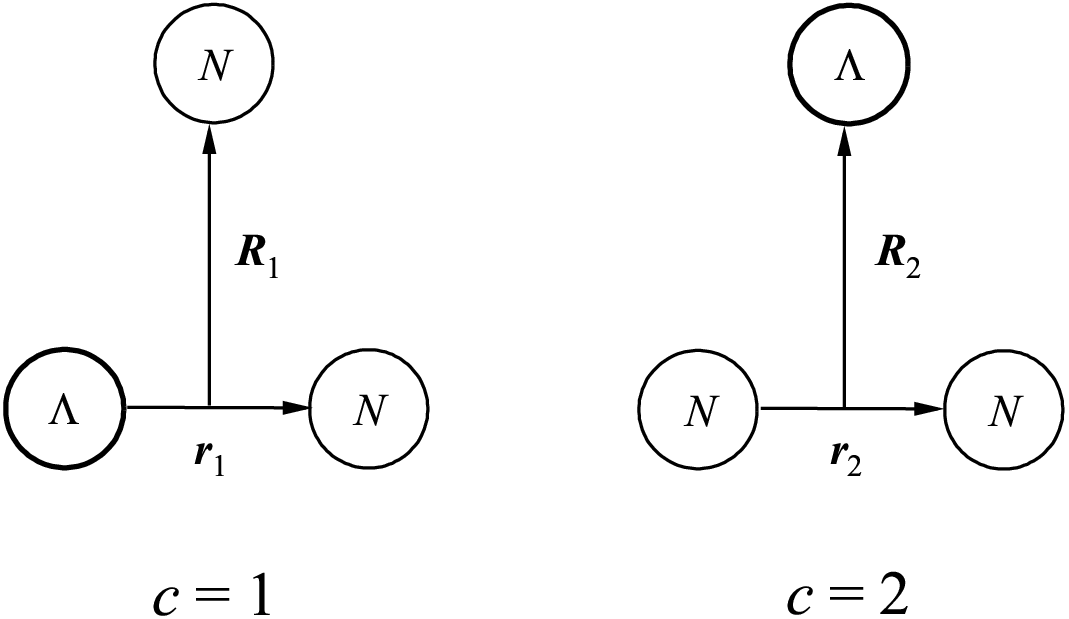}\
	\caption{Two sets of Jacobi coordinates of the $N N \Lambda$ three-body system.}
	\label{jacobian}
\end{figure}

To investigate the $NN\Lambda$ three-body system, we employ the GEM to solve the three-body Schr\"odinger equation.
Two sets of Jacobi coordinates are introduced to describe the relative motions among the three particles.
The total wave function corresponding to the two sets of Jacobi coordinates, labeled as $c$ = 1 and 2 in Fig.~\ref{jacobian}, can be written as
\begin{align}
	\Psi_{J M T T_{z}}\left(N N \Lambda\right) = & \sum_{c=1}^{2} \sum_{s, S, L} \sum_{n_{1}, l_{1}} \sum_{n_{2}, l_{2}} C_{\gamma}^{(c)} \mathcal{A} \nonumber \\ 
	& \times \left\{\left\{\left(\phi_{n_{1} l_{1}}^{(c)}\left(\boldsymbol{r}_{c}\right) \psi_{n_{2} l_{2}}^{(\mathrm{c})}\left(\boldsymbol{R}_{c}\right)\right)_{L}\right.\right. \nonumber \\
	& \left.\left. \times \left[\left(\chi_{\frac{1}{2}}^{1} \chi_{\frac{1}{2}}^{2}\right)_{s} \chi_{\frac{1}{2}}^{3}\right]_{S}\right\}_{J M}\left(\tau_{\frac{1}{2}}^{1} \tau_{\frac{1}{2}}^{2}\right)_{T T_{z}}\right\},
\end{align} 
where $\gamma$ denotes $\{L,s,S,n_1,l_1,n_2,l_2\}$.
Here, $\chi$ and $\tau$ represent the spin and isospin wave functions of the nucleons or the $\Lambda$ hyperon, respectively.
The operator $\mathcal{A}$ denotes the antisymmetrization operator for the two nucleons.
Within the GEM, the relative wave functions associated with the two Jacobi coordinates are expanded using Gaussian basis functions,
\begin{align}
	\phi_{n_{1} l_{1}}(\boldsymbol{r}) & =r^{\ell_{1}} e^{-\left(r / r_{n_{1}}\right)^{2}} Y_{\ell_{1} m_{1}}(\hat{\boldsymbol{r}}) \nonumber \\
	\psi_{n_{2} l_{2}}(\boldsymbol{R}) & =R^{\ell_{2}} e^{-\left(R / R_{n_{2}}\right)^{2}} Y_{\ell_{2} m_{2}}(\hat{\boldsymbol{R}}).
\end{align}

The Gaussian variational parameters are chosen to have geometric progression, allowing the basis to cover a wide range of length scales,
\begin{align}
	r_{n_{1}} &= r_{\min}A_{1}^{n_{1}-1}, ~~~~ \left(n_{1} = 1, 2, \ldots, n_{1}^{\max}\right), \nonumber \\
	R_{n_{2}} &= R_{\min}A_{2}^{n_{2}-1}, ~~~ \left(n_{2} = 1,2, \ldots, n_{2}^{\max}\right).
\end{align}
Then, the eigenenergies and the expansion coefficients $C_{\gamma}^{(c)}$ are obtained by applying the Rayleigh--Ritz variational method.
Further details of the GEM and its applications to few-body systems can be found in Refs.~\cite{Hiyama:2003cu,Hiyama:2019kpw,Wu:2023cfh}.

\section{Results and discussions}
\label{3}

In this section, we present the numerical results for the $N\Lambda$ interaction and explore its implications for the $p\Lambda$ femtoscopic correlation function and the $NN\Lambda$ three-body system.
We first analyze the bound-state and scattering properties of the $N\Lambda$ system, followed by applications of the resulting effective interaction to the two- and three-body systems.

\subsection{Bound-state and scattering analyses}
\label{31}

The bound-state calculations are carried out using the Resonating Group Method (RGM) and the Generator Coordinate Method (GCM).
The corresponding formalism is summarized in Appendix~\ref{appendix}, and further details can be found in Refs.~\cite{Wheeler:1937zza,Hill:1952jb,Griffin:1957zza}.
Table~\ref{energy} lists the coupled channels involved in the $N \Lambda$ systems with $J^P = 0^+$ and $1^+$, along with the corresponding lowest energies.
The coupled channels are $N \Lambda, N \Sigma, \Delta \Sigma^*$ for the $J^P = 0^+$ system, and $N \Lambda, N \Sigma, N \Sigma^*, \Delta \Sigma, \Delta \Sigma^*$ for the $J^P = 1^+$ system.
A bound state is identified when the lowest eigenenergy lies below the lowest two-body threshold.
Specifically, the lowest energy of the $N \Lambda$ system with $J^P = 0^+$ after channel coupling is found to be 2066 MeV, while that of the $N \Lambda$ system with $J^P = 1^+$ is 2067 MeV.
Since these energies are above the $N\Lambda$ threshold, the $N\Lambda$ systems with $J^P = 0^+$ and $1^+$ are not bound in the current calculation.
It is worth noting that for unbound systems, the present calculation is performed in a finite space with a limited number of basis functions. 
For an unbound system, the lowest discretized eigenenergy is expected to approach the corresponding threshold as the model space is enlarged.

\begin{table}
	\caption{Coupled channels involved in the $N \Lambda$ systems and the systems' lowest energies. The threshold of the $N \Lambda$ systems is 2063 MeV.}
	\begin{tabular}{c c c}
		\hline \hline
		~~~~$J^{P}$~~~~ & Coupled channels  & ~~~~Lowest energy~~~~  \\ \hline
		$0^{+}$  & $N \Lambda, N \Sigma, \Delta \Sigma^*$ & 2066 MeV  \\
		$1^{+}$  & $N \Lambda, N \Sigma, N \Sigma^*, \Delta \Sigma, \Delta \Sigma^*$ & 2067 MeV \\
		\hline \hline
		\label{energy}
	\end{tabular}
\end{table}

\begin{figure}[htb]
	\centering
	\includegraphics[width=8.6cm]{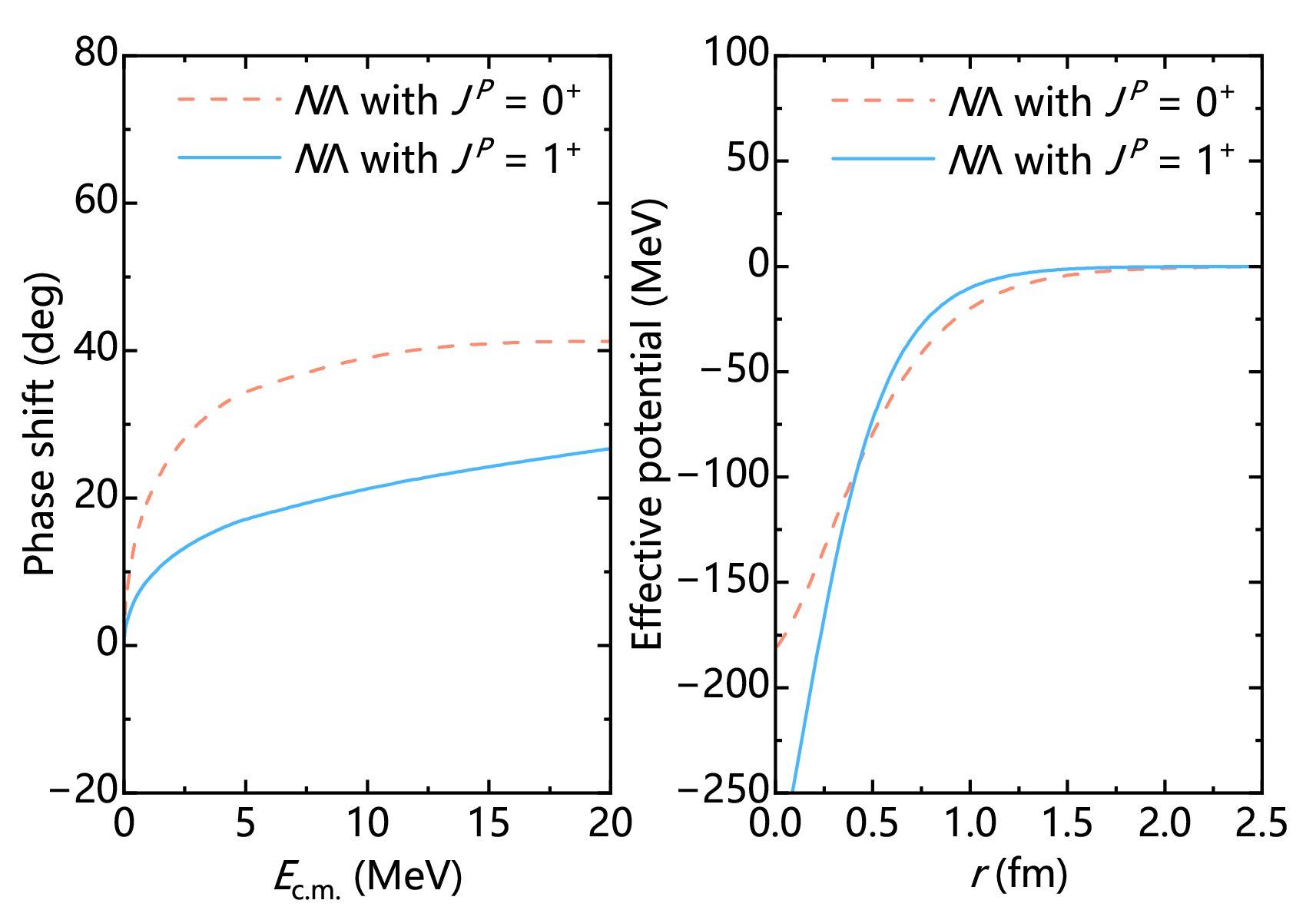}\
	\caption{Scattering phase shifts and effective potentials of the $N \Lambda$ system with $J^P = 0^+$ and $1^+$.}
	\label{scattering}
\end{figure}

We calculate the $N\Lambda$ scattering phase shifts using the extended Kohn--Hulth\'en--Kato (KHK) variational method.
The scattering formalism is summarized in Appendix~\ref{appendix}, and further details can be found in Ref.~\cite{Kamimura:1977okl}.
The left panel of Fig.~\ref{scattering} shows the scattering phase shifts of the $N\Lambda$ systems with channel coupling.
Therein, the dashed red line denotes the phase shift of the $N\Lambda$ system with $J^P = 0^+$, while the solid blue line corresponds to that with $J^P = 1^+$.
For both spin channels, the phase shifts are positive throughout the energy range considered, indicating an overall attractive $N\Lambda$ interaction.

In the $J^P = 0^+$ channel, the phase shift increases rapidly at low energies and saturates around $40^\circ$, implying a relatively strong attraction. 
In contrast, the $J^P=1^+$ channel exhibits a more moderate increase, indicating a weaker but still attractive interaction.
As the energy approaches the threshold, neither phase shift tends to $180^\circ$, as would be expected for an $S$-wave channel containing bound states according to Levinson's theorem.
This indicates that although the interaction is attractive, it is not strong enough to generate a bound state in the $N\Lambda$ system under the current model calculation.
These phase shift behaviors are consistent with the eigenenergy analysis presented earlier, in which the lowest eigenvalues of the $N\Lambda$ systems with $J^P = 0^+$ and $1^+$ remain above the $N\Lambda$ threshold after channel coupling. 
Based on the scattering phase shifts, the scattering parameters for the $N\Lambda$ system have been extracted and are listed in Table~\ref{scatteringparam}.
The scattering length and effective range are obtained as $a_0=-2.36$ fm and $r_{\rm eff}=1.61$ fm for $J^P=0^+$, and $a_0=-0.95$ fm and $r_{\rm eff}=1.46$ fm for $J^P=1^+$, respectively.

\begin{table}
	\caption{Scattering length and effective range for the $N\Lambda$ system.}
	\begin{tabular}{c c c}
		\hline \hline
		~~~~$J^{P}$~~~~ & ~~Scattering length~~  & ~~~~Effective range~~~~  \\ \hline
		$0^{+}$  & $-$2.36 fm & 1.61 fm  \\
		$1^{+}$  & $-$0.95 fm & 1.46 fm  \\
		\hline \hline
		\label{scatteringparam}
	\end{tabular}
\end{table}

Furthermore, effective $N\Lambda$ potentials are reconstructed from the scattering parameters obtained in the coupled-channel calculation using the GLM inverse-scattering method described in Sec.~\ref{23}.
The resulting effective potentials are displayed in the right panel of Fig.~\ref{scattering}.
As shown in the figure, the two spin channels exhibit different attractive behaviors in different radial regions.
Although the $J^P=1^+$ channel can be more attractive at shorter distances, the stronger attraction of the $J^P=0^+$ potential at larger distances is more relevant to the low-energy and near-threshold dynamics, and is therefore expected to play a more important role in the binding properties and the femtoscopic correlation function.
This difference can be primarily attributed to the spin dependence of the quark-quark interaction in the QDCSM.
In particular, the chromomagnetic interaction in Eq.~(\ref{Voge}) contributes differently to the two spin channels because of their different spin-color configurations.
The resulting spin-dependent interaction leads to the distinct radial behavior of the two effective potentials and is consistent with the stronger low-energy attraction indicated by the $J^P=0^+$ scattering phase shifts. 

\subsection{$p \Lambda$ femtoscopic correlation function}
\label{32}

In this part, we investigate the $p\Lambda$ femtoscopic correlation function based on the KP formula.
A static and spherically symmetric Gaussian source is adopted in the present calculation, and recent studies of the emission source can be seen in Refs.~\cite{ALICE:2020ibs,Mihaylov:2023pyl,ALICE:2023sjd,VazquezDoce:2024nye,Wang:2025htd,ALICE:2025,ALICE:2025zzg,Wang:2024bpl}.
The source size for the $p\Lambda$ system is constrained by studies of $pp$ and $p\Lambda$ pairs, which support a common core Gaussian source after accounting for contributions from strongly decaying resonances~\cite{ALICE:2020ibs,ALICE:2021njx,ALICE:2025}.
The core source radius depends on the transverse mass of the pair.
For $p\Lambda$ pairs with an average transverse mass of $\langle m_{\rm T}\rangle=1.55$ GeV, the originally reported values were $r_{\rm core}=1.02\pm0.04$ fm and an effective source size of $R=1.23$ fm~\cite{ALICE:2021njx,ALICE:2020ibs}.
These values were subsequently updated to $r_{\rm core}=0.94\pm0.04$ fm and $R=1.15\pm0.04$ fm~\cite{ALICE:2025}.
Accordingly, the updated value $R=1.15$ fm is adopted in the present calculation.

The relative wave function is obtained by solving the two-body Schr\"odinger equation for the $p\Lambda$ system.
Since the $\Lambda$ is electrically neutral, only the strong $N\Lambda$ interaction needs to be considered.
Based on the aforementioned analysis, the effective $N\Lambda$ interaction is reconstructed from the scattering parameters obtained in the coupled-channel calculation, thereby incorporating the corresponding channel-coupling effects.
A similar effective single-channel treatment of coupled-channel effects has also been employed in studies of the $\Xi N$ and $\Xi\alpha$ interactions and their femtoscopic correlations~\cite{Hiyama:2019kpw,Hiyama:2022jqh,Kamiya:2024diw}.

\begin{figure}[htb]
	\centering
	\includegraphics[width=7.6cm]{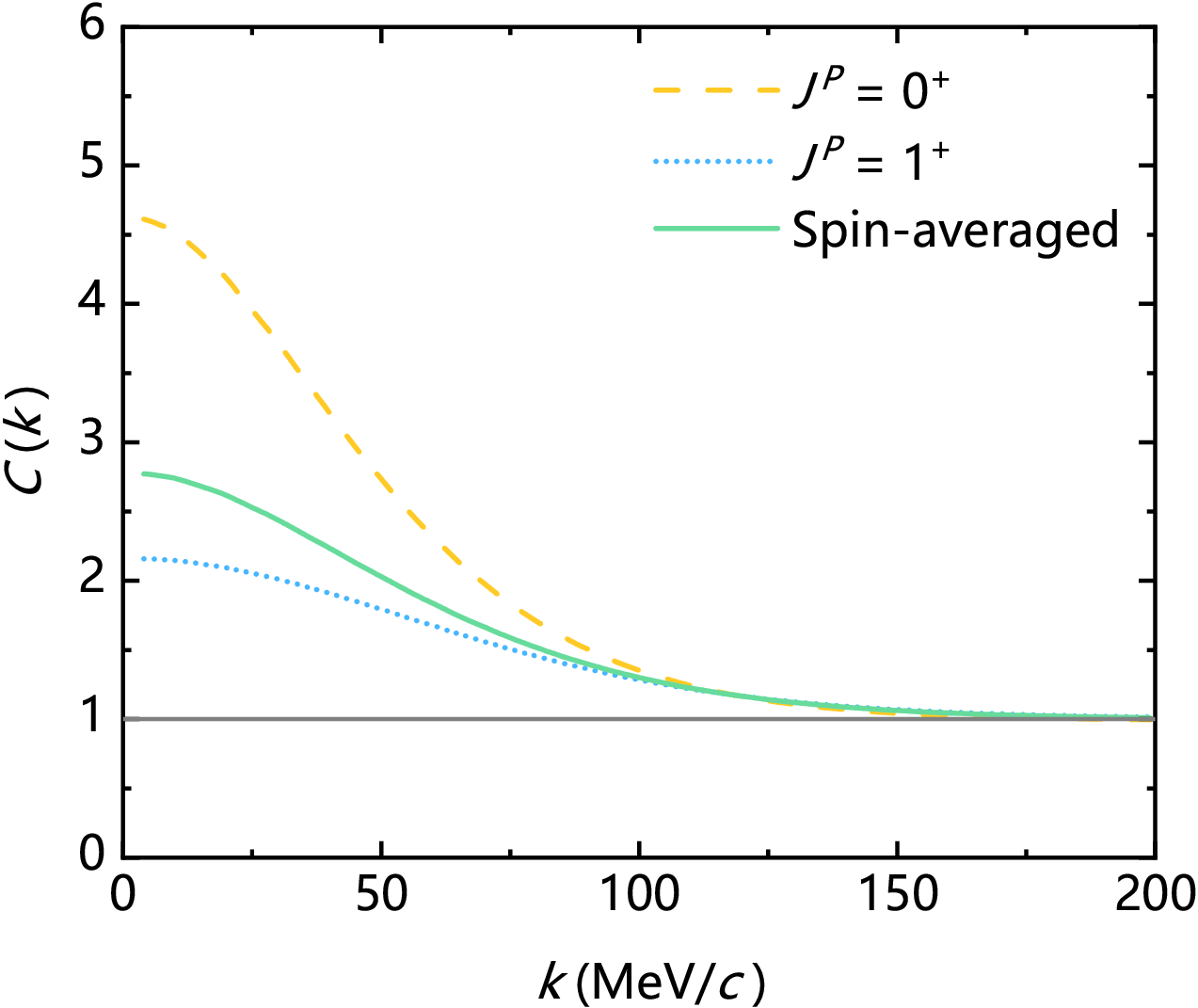}\
	\caption{$p \Lambda$ correlation function calculated based on the QDCSM.}
	\label{CF}
\end{figure}

As shown in Fig.~\ref{CF}, the $p\Lambda$ correlation function is calculated with the source size parameter $R = 1.15$ fm.
The dashed yellow line represents the result for the $p \Lambda$ with $J^P = 0^+$, the dotted blue line corresponds to $J^P = 1^+$, and the solid green line denotes the spin-averaged result according to Eq.~\ref{average}.
The $p\Lambda$ correlation function with $J^P=0^+$ exhibits a considerably stronger enhancement than that for $J^P=1^+$, consistent with the stronger low-energy attraction in the singlet channel.
For both spin channels, the calculated correlation functions remain above or close to unity over the momentum range considered.
The spin-averaged result lies between the two individual components, as expected from the statistical weights of the singlet and triplet channels.

For comparison with the experimental data, residual correlations arising from feed-down contributions must also be taken into account.
The dominant contributions considered here originate from the decays $\Sigma^0\to\Lambda\gamma$, $\Xi^0\to\Lambda\pi^0$, and $\Xi^-\to\Lambda\pi^-$.
These processes induce residual correlations involving $p\Sigma^0$, $p\Xi^0$, and $p\Xi^-$ pairs. 
Consequently, the residual feed-down correlation of $\Lambda$ hyperons originating from the decay of a $\Sigma^0$ is denoted as $C_{p(\Sigma^{0})}(k)$, while that arising from $\Xi$ decays into $\Lambda$ is denoted as $C_{p(\Xi)}(k)$.
The residual $p\Sigma$ and $p\Xi$ correlation functions are evaluated using interactions taken from chiral EFT~\cite{Haidenbauer:2013oca} and lattice QCD-based
calculations~\cite{HALQCD:2019wsz,ALICE:2019hdt}, respectively.
Using chiral EFT and lattice QCD inputs for the residual interactions, as in Refs.~\cite{ALICE:2021njx,Garrido:2024pwi}, facilitates a comparison of different $p\Lambda$ interactions by reducing differences associated with the feed-down modeling.
Other feed-down contributions are subdominant and can be approximated by a constant term $C_\text{flat} \approx 1$. 
The contribution associated with misidentified protons carries a weight of only about $0.6\%$ in the ALICE analysis and is approximated as a flat contribution; therefore, it is not treated explicitly here.

The corrected $p\Lambda$ correlation function is then expressed as a weighted sum of the primary and residual components, along with the flat background term, as follows:
\begin{align}
	C_{\text {corrected}}(k)=& \ B(k)\left[\lambda_{p \Lambda} C_{p \Lambda}(k)+\lambda_{p(\Sigma^{0})} C_{p(\Sigma^{0})}(k)\right. \nonumber \\
	& \left.+\lambda_{p(\Xi)} C_{p(\Xi)}(k)+\lambda_{\mathrm{flat}} \right].
\end{align}
Here, $\lambda_{\mathrm{flat}}$ effectively includes the subdominant flat contributions, including the small contamination from misidentified protons.
$B(k)$ accounts for the non-femtoscopic baseline contribution.
In the low-momentum region considered here, it is taken as a constant normalization factor determined from the fit to the experimental correlation function.
The residual contributions from $p\Sigma$ and $p\Xi$ pairs need to be transformed into the momentum basis of the $p\Lambda$ correlation.
This transformation is performed following the prescription of Ref.~\cite{Kisiel:2014mma}.
The weight parameters $\lambda$ are taken from the ALICE experimental analysis, where they are determined from the particle purities and the corresponding primary and secondary fractions.
In the present calculation, the non-flat component weights are varied within their reported systematic ranges together with the baseline normalization $B$, while $\lambda_{\mathrm{flat}}$ is determined by requiring the component fractions to sum to unity.
The values providing the best description of the data are adopted and summarized in Table~\ref{weight}.

\begin{table}
	\caption{Weight parameters of the different components entering the corrected $p\Lambda$ correlation function.}
	\begin{tabular}{c c c c c}
		\hline \hline
		& ~~~~$\lambda_{p \Lambda}$~~~~   & ~~~~$\lambda_{p(\Sigma^{0})}$~~~~ & ~~~~$\lambda_{p(\Xi)}$~~~~ & ~~~~$\lambda_{\mathrm{flat}}$~~~~   \\ \hline
		Ref.~\cite{ALICE:2021njx}   & $0.471^{+0.025}_{-0.044}$ & $0.157^{+0.023}_{-0.031}$  & $0.190^{+0.031}$  & 0.176  \\
		Ref.~\cite{Garrido:2024pwi} & 0.427 & 0.170  & 0.221  & 0.182  \\
		This work                   & 0.496 & 0.180  & 0.221  & 0.103  \\
		\hline \hline
		\label{weight}
	\end{tabular}
\end{table}

\begin{figure}[htb]
	\centering
	\includegraphics[width=8cm]{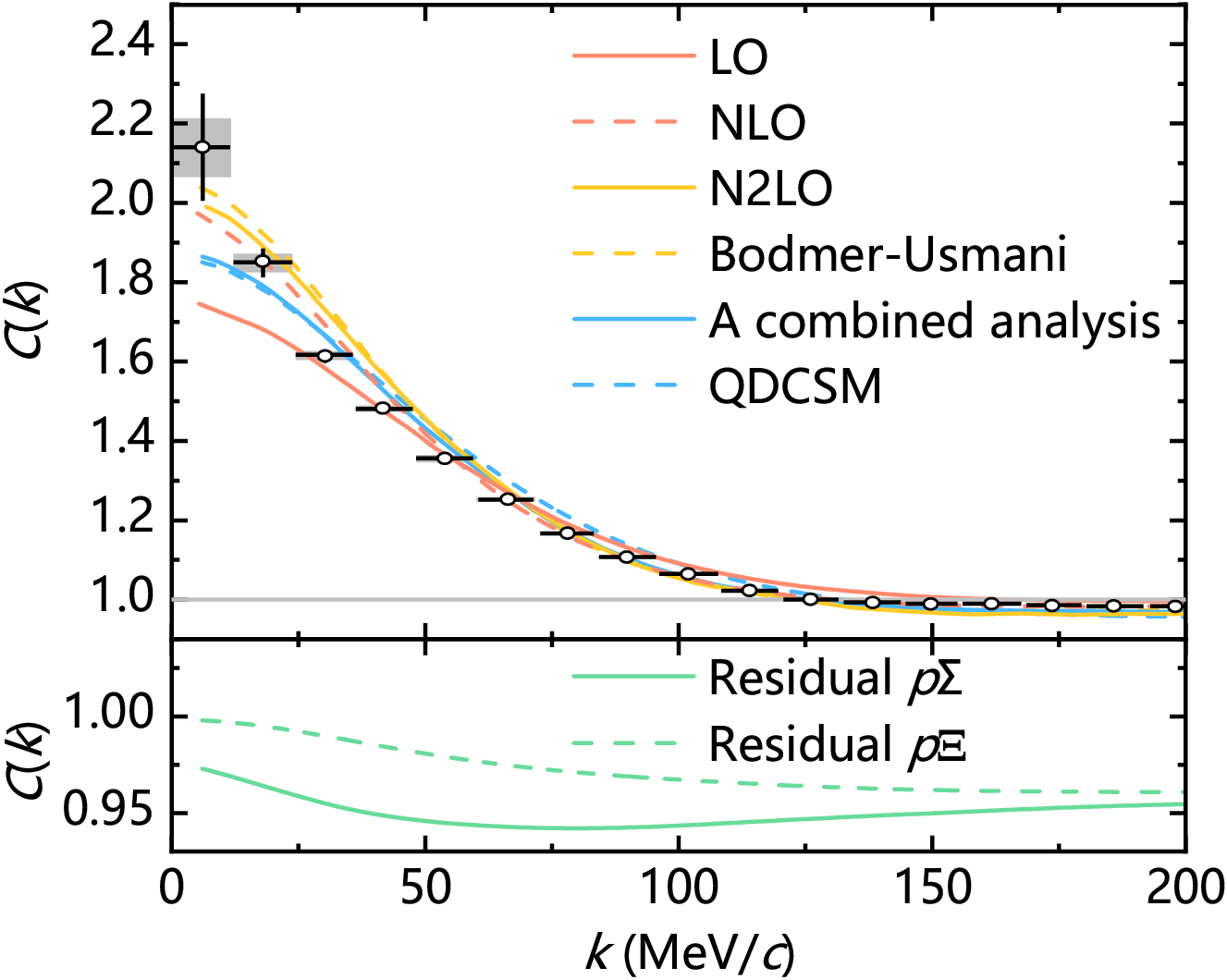}\
	\caption{Corrected $p \Lambda$ correlation functions calculated within different frameworks and the experimental measurements.}
	\label{CFs}
\end{figure}

Fig.~\ref{CFs} shows the experimental measurements~\cite{ALICE:2021njx}, including statistical (vertical bars) and systematic (grey boxes) uncertainties, together with the corrected $p\Lambda$ correlation functions calculated using different theoretical interactions.
The solid and dashed red lines represent the results obtained using the leading-order (LO)~\cite{Polinder:2006zh} and next-to-leading-order (NLO)~\cite{Haidenbauer:2013oca} chiral EFT interactions, respectively.
The solid and dashed yellow lines represent the results obtained in Ref.~\cite{Garrido:2024pwi} using an interaction based on chiral EFT at next-to-next-to-leading order (N2LO) and the Bodmer--Usmani $p\Lambda$ potential~\cite{Bodmer:1984gc,Bodmer:1988en}, respectively.
The solid blue line represents the result obtained using a $p\Lambda$ interaction constrained by a combined analysis of femtoscopic and scattering data~\cite{Mihaylov:2023ahn}.
The dashed blue line shows the result obtained using the $p\Lambda$ interaction constructed in the present work.
The solid and dashed green lines in the lower panel show the residual effects through $\mathcal{R}_{F}(k) = B(k) [ \lambda_{p(F)} C_{p(F)}(k) - \lambda_{p(F)}  +1]$ for $F=\Sigma^0$ and $\Xi$, respectively. 
Their deviations from $B(k)$ quantify the weighted contributions of the corresponding feed-down channels.

As shown in Fig.~\ref{CFs}, the QDCSM result underestimates the experimental correlation strength in the region $k<20$ MeV/$c$.
This discrepancy can be largely understood in terms of the relatively weak attraction in the $J=1$ $N\Lambda$ channel predicted by the QDCSM.
Since the triplet channel carries a statistical weight of $3/4$ in the spin-averaged correlation function, its interaction has a substantial influence on the correlation strength at low relative momentum.
In particular, Ref.~\cite{Mihaylov:2023ahn} provides a combined analysis of low-energy $p\Lambda$ scattering, considering both cross-section and correlation data, and extracts the scattering length of the $p\Lambda$ spin triplet to be $-1.56$ to $-1.18$ fm. 
However, the corresponding value obtained in the QDCSM is $a_0=-0.95$ fm for the $J^P=1^+$ channel, suggesting that the triplet attraction is underestimated in the present calculation.
The remaining differences among the theoretical calculations and the experimental data indicate that further constraints on the low-energy $N\Lambda$ interaction are still desirable.

\subsection{$N N \Lambda$ three-body bound state calculation}
\label{33}

The solution of the $NN\Lambda$ three-body Schr\"odinger equation requires both the nucleon--nucleon and nucleon--$\Lambda$ interactions.
The effective $N\Lambda$ potentials used here are those obtained from the inverse-scattering analysis discussed in Sec.~\ref{31}.
For the $NN$ interaction, the same inverse-scattering strategy is employed.
Within the framework of the QDCSM, the $NN$ scattering process has been investigated in previous work~\cite{Ping:2008tp}.
The deuteron is strongly influenced by the $D$-wave component, and therefore the coupled $S$-$D$ channel is taken into account in the $NN$ scattering calculation.
Two parameter sets for the $NN$ interaction are considered to examine the role of the $N\Lambda$ interaction in the $NN\Lambda$ three-body system under different $NN$ inputs.
The resulting low-energy parameters are $a_{NN}^{(1)}=5.94$ fm and $r_{NN}^{(1)}=1.75$ fm for the first parameter set, and $a_{NN}^{(2)}=6.03$ fm and $r_{NN}^{(2)}=1.67$ fm for the second.
Together with the corresponding deuteron bound-state information, these scattering parameters are used as inputs to the inverse-scattering procedure.
The effective $NN$ potentials are constructed using the same inverse-scattering procedure as that employed for the $N\Lambda$ interaction, with the effects of the $D$-wave coupling effectively incorporated into the resulting single-channel interactions.

\begin{figure}[htb]
	\centering
	\includegraphics[width=7.5cm]{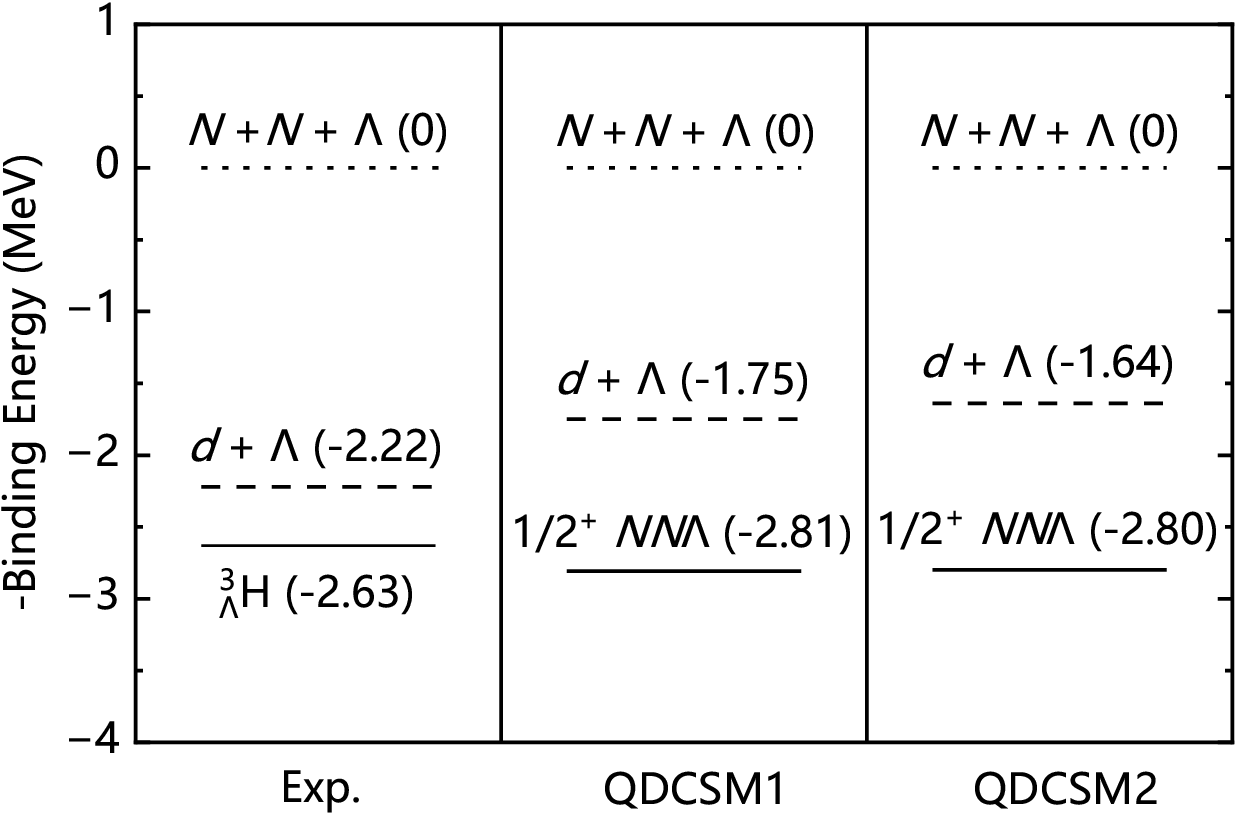}\
	\caption{Calculated binding energies of $^3_\Lambda\mathrm{H}$ relative to the $NN\Lambda$ three-body breakup threshold.
		The STAR measurement is compared with the theoretical results from the QDCSM, and the position of the $d+\Lambda$ threshold is also indicated.}
	\label{NNL}
\end{figure}

Fig.~\ref{NNL} displays the binding energies of the $^3_\Lambda \mathrm{H}$ hypernucleus relative to the $NN\Lambda$ three-body breakup threshold, where the $d + \Lambda$ threshold is also indicated. 
Several experimental determinations of the hypertriton binding energy are available in Refs.~\cite{Gajewski:1967ruj,Juric:1973zq,STAR:2019wjm,ALICE:2022sco}.
For comparison with the calculated total three-body binding energies, the STAR result is shown in Fig.~\ref{NNL}.
According to our calculation, the $NN\Lambda$ system with $J^P = 1/2^+$ forms a bound state, corresponding to $^3_\Lambda \text{H}$. 
The corresponding $\Lambda$ separation energy, $B_\Lambda = B(^3_\Lambda{\rm H})-B_d$, is found to be 1.06--1.16 MeV, while it is $0.41 \pm 0.12 \pm 0.11$ MeV in the STAR Collaboration measurement~\cite{STAR:2019wjm}.
The $NN\Lambda$ system with $J^P = 3/2^+$ is unbound, and therefore no corresponding binding energy is shown in the figure.

To further characterize the spatial structure of the calculated $J^P=1/2^+$ state, we evaluate the root-mean-square (RMS) separations between its constituent baryons using the full three-body wave function. 
For the first $NN$ parameter set, with a scattering length of 5.94 fm, the RMS $NN$ and $N\Lambda$ separations are approximately 3.1 and 4.5 fm, respectively. 
For the second set, with a scattering length of 6.03 fm, the corresponding separations are approximately 3.0 and 4.3 fm. 
The two nucleon--$\Lambda$ pair separations are identical owing to nucleon exchange symmetry in the equal-mass approximation adopted here. 
Both calculations yield a larger $N\Lambda$ separation than the $NN$ separation. 
Together with the spin-triplet character of the $NN$ subsystem considered here, this spatial hierarchy is consistent with a deuteron-like $NN$ core coupled to a more extended $\Lambda$ distribution.

In the $NN\Lambda$ three-body system, both singlet and triplet components of the $N\Lambda$ interaction are involved. 
Owing to the deuteron-like $NN$ subsystem with $S_{NN}=1$, the $\Lambda$ hyperon can couple to the two nucleons through both spin channels. 
For the $J^P=1/2^+$ $NN\Lambda$ state, both singlet and triplet $N\Lambda$ components contribute, with the singlet interaction playing a more important role in generating the binding.
By contrast, the $J^P=3/2^+$ configuration is governed predominantly by the triplet $N\Lambda$ interaction.
The stronger attraction in the singlet $N\Lambda$ channel therefore favors binding in the $J^P=1/2^+$ system, whereas the weaker triplet interaction is insufficient to bind the $J^P=3/2^+$ state.
Although the present calculation predicts a larger $\Lambda$ separation energy than the experimental values, it reproduces the weakly bound nature of the hypertriton.
The $J^P=3/2^+$ state is predicted to remain unbound in the present calculation.
The overall qualitative description supports the extension of the present framework to other three-body hadronic systems, such as the $NN\bar{K}$ system~\cite{Shevchenko:2006xy,Dote:2008hw,Ikeda:2010tk,Bayar:2013lxa,Kezerashvili:2016ztc,Dote:2017wkk,MartinezTorres:2020hus}, in future studies.

\section{Summary}

In this work, the $N\Lambda$ interaction has been systematically investigated within the framework of the quark delocalization color screening model.
The energy calculations show that the $N\Lambda$ system does not support a bound state in either the $J^P=0^+$ or $1^+$ channel.
The scattering phase shifts indicate weak attraction in both channels, with a stronger low-energy attraction in the $J^P=0^+$ channel.
The corresponding scattering lengths and effective ranges are obtained as $a_0=-2.36$ fm and $r_{\rm eff}=1.61$ fm for the singlet channel, and $a_0=-0.95$ fm and $r_{\rm eff}=1.46$ fm for the triplet channel.

Using the GLM inverse-scattering method, effective $N\Lambda$ potentials are reconstructed from the low-energy scattering information obtained in the coupled-channel calculation.
The resulting interactions are applied to the $p\Lambda$ femtoscopic correlation function, where the $J^P=0^+$ channel produces a stronger enhancement than the $J^P=1^+$ channel, consistent with its stronger low-energy attraction.
The spin-averaged QDCSM result underestimates the experimental correlation strength at very low relative momentum, which can be mainly attributed to the relatively weak triplet $N\Lambda$ interaction predicted by the model.

The $NN\Lambda$ three-body system is further studied using the GEM.
The $J^P=1/2^+$ hypertriton is found to be weakly bound, with a $\Lambda$ separation energy of $1.06$--$1.16$ MeV, while the $J^P=3/2^+$ state remains unbound.
Although the calculated $\Lambda$ separation energy is larger than the experimental values, the present framework reproduces the qualitative weak-binding feature of the hypertriton.
These results provide complementary constraints on the singlet and triplet components of the low-energy $N\Lambda$ interaction and support the extension of the present framework to other hypernuclear and three-body hadronic systems.
	
\acknowledgments{This work is supported partly by the National Natural Science Foundation of China under Contracts Nos. 12305087 and 12575088. 	
	Y.Y. is supported by the Scientific Research Foundation of Changzhou University of Information Technology under Grant No. SG210201B13003 and by the Changzhou Sci\&Tech Program under Grant No. 20260098.
	Q. W. is supported by the Natural Science Foundation of Jiangsu Province under grant No. BK20220122, the China Postdoctoral Science Foundation under grant No. 2024M751369, and the Jiangsu Funding Program for Excellent Postdoctoral Talent.}	

\setcounter{equation}{0}
\renewcommand\theequation{A\arabic{equation}}

\appendix

\section{RGM treatment of the bound-state and scattering processes}
\label{appendix}

The RGM is employed to investigate both the bound-state properties and the scattering processes of the $N\Lambda$ system.
For a two-baryon system, the total wave function can be written as
\begin{align}
	\Psi=\mathcal{A}\left[\Phi_{A}\Phi_{B}\chi(\boldsymbol{R})\right],
\end{align}
where $\Phi_{A}$ and $\Phi_{B}$ denote the internal wave functions of the two baryon clusters, $\chi(\boldsymbol{R})$ is their relative-motion wave function, and $\mathcal{A}$ is the antisymmetrization operator acting on identical quarks.

By applying the variational principle to the relative-motion wave function, the RGM equation is obtained as
\begin{align}
	\int \left[H(\boldsymbol{R},\boldsymbol{R}')-E N(\boldsymbol{R},\boldsymbol{R}')\right]\chi(\boldsymbol{R}')\,d\boldsymbol{R}'=0,
\end{align}
where $H(\boldsymbol{R},\boldsymbol{R}')$ and $N(\boldsymbol{R},\boldsymbol{R}')$ are the Hamiltonian and normalization kernels, respectively.
The relative-motion wave function is expanded in a set of generator-coordinate Gaussian basis functions,
\begin{align}
	\chi_{L}(\boldsymbol{R})=\sum_{i}C_{iL}\,\Gamma_{iL}(\boldsymbol{R},\boldsymbol{S}_{i}),
\end{align}
where $\boldsymbol{S}_{i}$ denotes the generator coordinate and $L$ is the orbital angular momentum between the two baryon clusters.
With this expansion, the RGM equation is transformed into the generalized eigenvalue equation
\begin{align}
	\sum_{j}H_{ij}C_{j}=E\sum_{j}N_{ij}C_{j},
\end{align}
where $H_{ij}$ and $N_{ij}$ are the Hamiltonian and overlap matrix elements, respectively.
The eigenenergies and the corresponding wave functions are obtained by solving this generalized eigenvalue problem.

For the scattering calculation, the relative-motion wave function is constructed according to the extended KHK variational method.
For a given partial wave $L$, it is expanded as
\begin{align}
	\chi_{L}(\boldsymbol{R})=\sum_{i}C_{i}\frac{\widetilde{u}_{L}(R,S_{i})}{R}Y_{LM}(\hat{\boldsymbol{R}}),
\end{align}
where the radial basis function $\widetilde{u}_{L}(R,S_{i})$ is defined by
\begin{align}
	\widetilde{u}_{L}(R,S_{i})=
	\begin{cases}
		\alpha_{i}u_{L}(R,S_{i}), & R\leq R_{C},\\
		\hat{h}_{L}^{(-)}(kR)-s_{i}\hat{h}_{L}^{(+)}(kR), & R\geq R_{C},
	\end{cases}
\end{align}
where $R_{C}$ is the matching radius beyond which the strong interaction can be neglected, $k$ is the relative momentum, and $\hat{h}_{L}^{(\pm)}$ denote the incoming and outgoing Riccati--Hankel functions.
The parameters $\alpha_i$ and $s_i$ are determined by requiring the radial wave function and its derivative to be continuous at $R=R_C$.

The expansion coefficients $C_i$ are determined variationally.
Following the KHK procedure, one obtains the linear equations
\begin{align}
	\sum_{j=1}^{n-1}L_{ij}^{L}C_{j}=M_{i}^{L}, \qquad i=1,2,\ldots,n-1,
\end{align}
with
\begin{align}
	L_{ij}^{L}&=K_{ij}^{L}-K_{i0}^{L}-K_{0j}^{L}+K_{00}^{L}, \\
	M_{i}^{L}&=K_{00}^{L}-K_{i0}^{L},
\end{align}
where
\begin{align}
	K_{ij}^{L}=\left\langle \widehat{\Phi}_{i}^{L}\left|H-E\right|\widehat{\Phi}_{j}^{L}\right\rangle .
\end{align}
Here, $\widehat{\Phi}_{i}^{L}$ represents the antisymmetrized channel wave function containing the corresponding radial basis function.

After solving the linear equations, the partial-wave $S$-matrix element is obtained as
\begin{align}
	S_{L}=\sum_{i}C_{i}s_{i}.
\end{align}
For elastic scattering, the $S$-matrix can be expressed as
\begin{align}
	S_{L}=e^{2i\delta_{L}},
\end{align}
from which the scattering phase shift $\delta_{L}$ is extracted.

For coupled-channel scattering, the channel indices are introduced into the wave functions and the corresponding variational equations, and all coupled channels are solved simultaneously.
The elastic $N\Lambda$ $S$-matrix element obtained in this way contains the effects of the coupling to the other channels and is used to determine the $N\Lambda$ scattering phase shifts presented in the main text.
Further details of the RGM and the extended KHK variational method can be found in Refs.~\cite{Wheeler:1937zza,Kamimura:1977okl,Hill:1952jb,Griffin:1957zza}.

\end{document}